\documentclass[seceq]{ptptex}

\usepackage{graphicx}

\def\lsim{ \,\, \vcenter{\hbox{$\buildrel{\displaystyle <}\over\sim$}}
 \,\,}
\def\be{\begin{equation}}
\def\ee{\end{equation}}
\def\bea{\begin{eqnarray}}
\def\eea{\end{eqnarray}}

\preprintnumber[4cm]{
BCCUNY-HEP/10-04, RBRC-864} 

\title{
Quarkonium in a non-ideal hot QCD Plasma
}

\author{
Adrian \textsc{Dumitru}%
}

\inst{
Department of Natural Sciences, Baruch College, CUNY,
17 Lexington Avenue, New York, NY 10010, USA\\
RIKEN BNL Research Center, Brookhaven National
Laboratory, Upton, NY 11973, USA
}

\abst{Substantial anisotropies should occur in the hot expanding QCD
  plasma produced in relativistic heavy-ion collisions due to
  non-vanishing shear viscosity. We discuss the form of the real-time,
  hard thermal loop resummed propagator for static gluons in the
  presence of such anisotropies and the consequences for quarkonium
  binding. It has been predicted that the propagator develops an
  imaginary part due to Landau damping at high temperature. This
  should generate a much larger width of quarkonium states than the
  Appelquist-Politzer vacuum estimate corresponding to decay into
  three gluons. We argue that this might be observable in heavy-ion
  collisions as a suppression of the $\Upsilon(1S)\to e^+ e^-$
  process. Lastly, we consider the heavy quark (singlet) free energy
  just above the deconfinement temperature. In the ``semi-QGP'',
  $F_{Q\bar{Q}}(R)$ at distances beyond $1/T$ is expected to be
  suppressed by $1/N$ as compared to an ideal plasma.}

\PTPindex{231}  

\begin{document}

\maketitle

\section{The anisotropic QGP}
This section deals with kinetic non-equilibrium effects in an
expanding QCD plasma. During the first few fm/c when the temperature
is high it is the longitudinal Bjorken expansion~\cite{Bjorken:1982qr}
which matters most; for semi-peripheral collisions and/or close to the
periphery of the fireball the transverse expansion may also be
important but is neglected here for simplicity. As a consequence of
the expansion, the particle momentum distribution in the local rest
frame is anisotropic, if the scattering rate is finite: there is a net
loss of particles with large $|p_z|$ from the fluid cell and
redistribution of the momenta requires time.  We assume that the
momentum distribution can be parameterized as
follows~\cite{Martinez:2010sc}:
\begin{equation}
f({\bf p}) = f_{\rm iso}(\sqrt{{\bf p}^2 + \xi ({\bf p}\cdot{\bf
    n})^2})  \simeq
f_{\rm iso}(p) \left[ 1-\xi \frac{({\bf p}\cdot{\bf n})^2}
{2pT} \left( 1\pm f_{\rm iso}(p)\right)
\right]~.  \label{eq:f_aniso}
\end{equation}
The second expression is valid to first order in the anisotropy
parameter $\xi$. $f_{\rm iso}(p)$ is either a Bose or a Fermi
distribution, respectively, and ${\bf n} = {\bf e_z}$ for longitudinal
expansion. Such a correction $\delta f$ to the equilibrium
distribution is also commonly employed in viscous hydrodynamics and
has been argued to be consistent with azimuthal flow coefficients
observed at RHIC~\cite{Lacey:2010fe} (a transverse unit vector ${\bf
  n} = {\bf e_T}$ is used for calculating $p_T$ distributions). Near
equilibrium one can relate moments of $\delta f$, which are
proportional to $\xi$, to the shear~\cite{Martinez:2010sc}.

\begin{figure}[htb]
\begin{center}
\includegraphics[width=7cm]{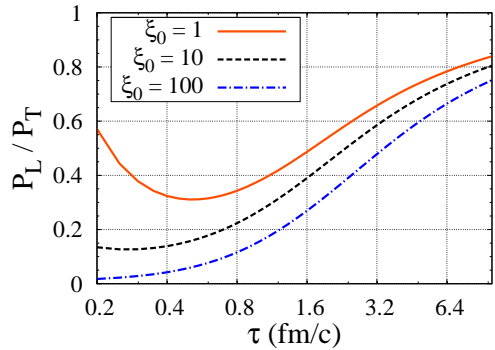}
\end{center}
\caption[a]{Pressure anisotropy during the high-temperature stage of a
  heavy-ion collision~\cite{Martinez:2010sc} for a
  low-viscosity plasma with $\eta/s=0.1$. The curves
  correspond to different {\em initial} anisotropies at $\tau_0=0.2$~fm/c.}
\label{fig:Paniso}
\end{figure}
Fig.~\ref{fig:Paniso} shows the time evolution of the ratio of the
longitudinal to the transverse pressure in the central region of a
high-energy heavy-ion collision (see, also,
ref.~\cite{Florkowski:2010cf}). This was obtained in
ref.~\cite{Martinez:2010sc} from a Boltzmann equation in relaxation
time approximation to all orders in the anisotropy parameter
$\xi$. Near equilibrium, the assumed equilibration rate translates
into a shear viscosity to entropy density ratio of
$\eta/s=0.1$. However, even for such ``strong coupling'' conditions
large pressure anisotropies are caused by the rapid longitudinal
expansion during the early, high-temperature stage of a heavy-ion
collision ($\tau \lsim 3$~fm/c). Note that $p_L<p_T$ does not suppress
transverse hydrodynamic flow~\cite{Heinz:2002rs} but may reflect in
the system-size and energy dependence of the transverse energy
$dE_T/dy$ in the final state~\cite{Dumitru:2001kz}.

\subsection{Real part of the quarkonium potential}
Anisotropies of the momentum distributions affect various processes in
the plasma. In particular, rotational symmetry in the local rest frame
is broken and the screening length acquires an angular dependence. At
high temperature and in the weak-coupling approximation it can be
obtained explicitly from the ``hard thermal loop'' (HTL) resummed
propagator for static electric gluons. To linear order in
$\xi$~\cite{Dumitru:2007hy},
\bea
{\rm Re}~\Delta^{00}({\bf p}) &=& \frac{1}{p^2+m_D^2} \left( 1 -
 \xi m_D^2 \frac{\frac{2}{3}-({\bf p}\cdot{\bf n})^2/p^2}{p^2+m_D^2} \right)~.
\eea
Here, $m_D=gT\sqrt{N_c/3}$ denotes the standard Debye mass of the
equilibrium plasma. The one-gluon exchange potential
follows essentially from the Fourier transform,
\begin{eqnarray} \label{eq:anisoPotlin_xi}
{\rm Re}~V({\bf{r}}) &=& {V}_{\rm iso}(r)
\left(1+\xi \left[\frac{\hat{r}}{6}+\frac{\hat{r}^2}{48}
+\frac{\hat{r}^2}{16}\cos(2\theta)\right] \right)~.
\end{eqnarray} 
Here, $\hat{r}\equiv r\,m_D$, $\cos\theta\equiv {\bf r}\cdot{\bf
  n}/r$, and ${V}_{\rm iso}(r)=-\frac{\alpha_s C_F}{r}\exp(-\hat{r})$
is the well-known Debye-screened Coulomb potential. The
potential~(\ref{eq:anisoPotlin_xi}) is valid for distances
$\hat{r}\lsim 1$. At fixed $T$ the $\xi$-dependent correction in
eq.~(\ref{eq:anisoPotlin_xi}) reduces thermal screening effects as
compared to an ideal plasma in local equilibrium ($\xi=0$). On the
other hand, one can absorb this $\xi$ dependence to a large extent by
a redefinition of the hard scale $T(\xi)$ and/or of the Debye mass
$m_D(\xi)$~\cite{Burnier:2009yu}. The most useful approach with
respect to applications to heavy-ion collisions would probably amount
to adjusting the initial $T_0$ in such a way that the entropy (per
unit of rapidity) in the final state remains fixed as $\xi_0$ is
varied. The resulting implicit dependence $T_0(\xi_0)$ can be
determined only from solutions of viscous hydrodynamics or transport
theory (with gluon-number changing processes~\cite{Baier:2000sb}).

\begin{figure}[hbt]
\begin{center}
\includegraphics[width=6.5cm]{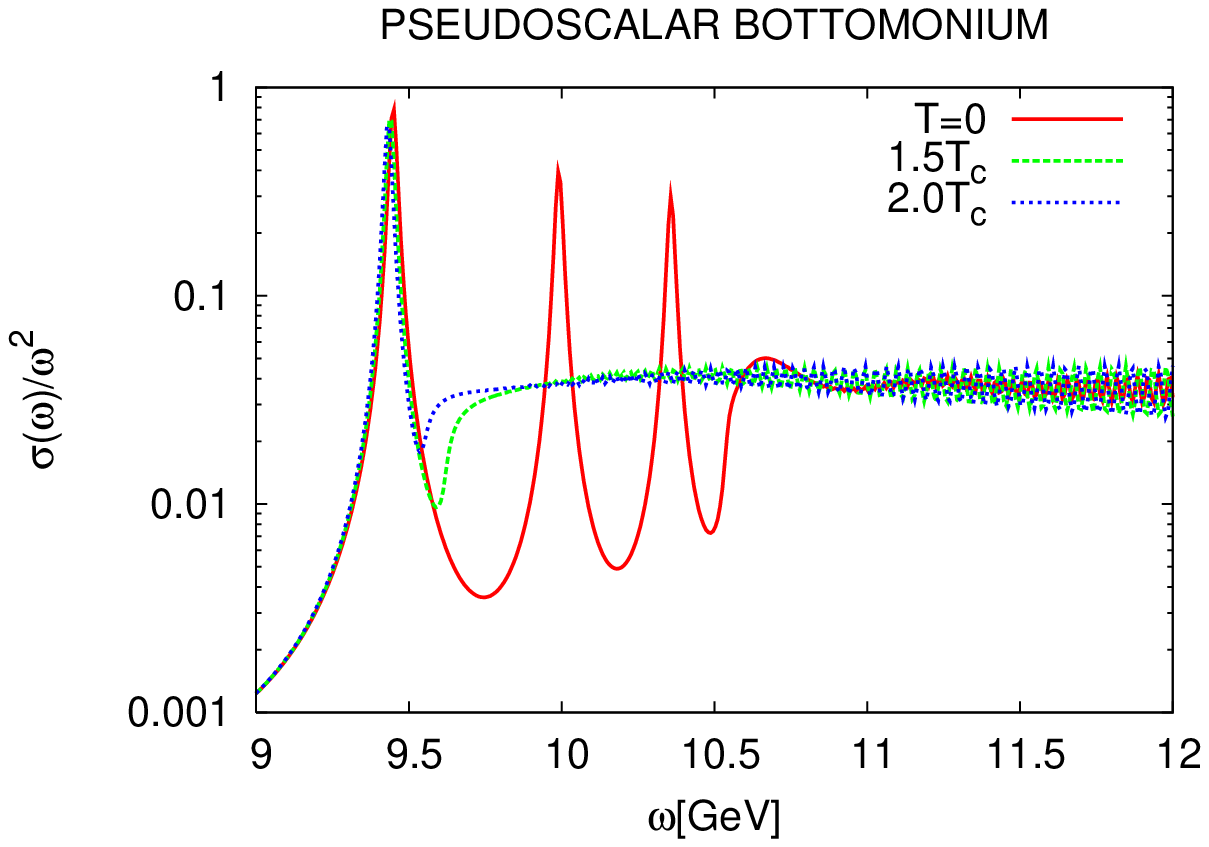}
\includegraphics[width=6.5cm]{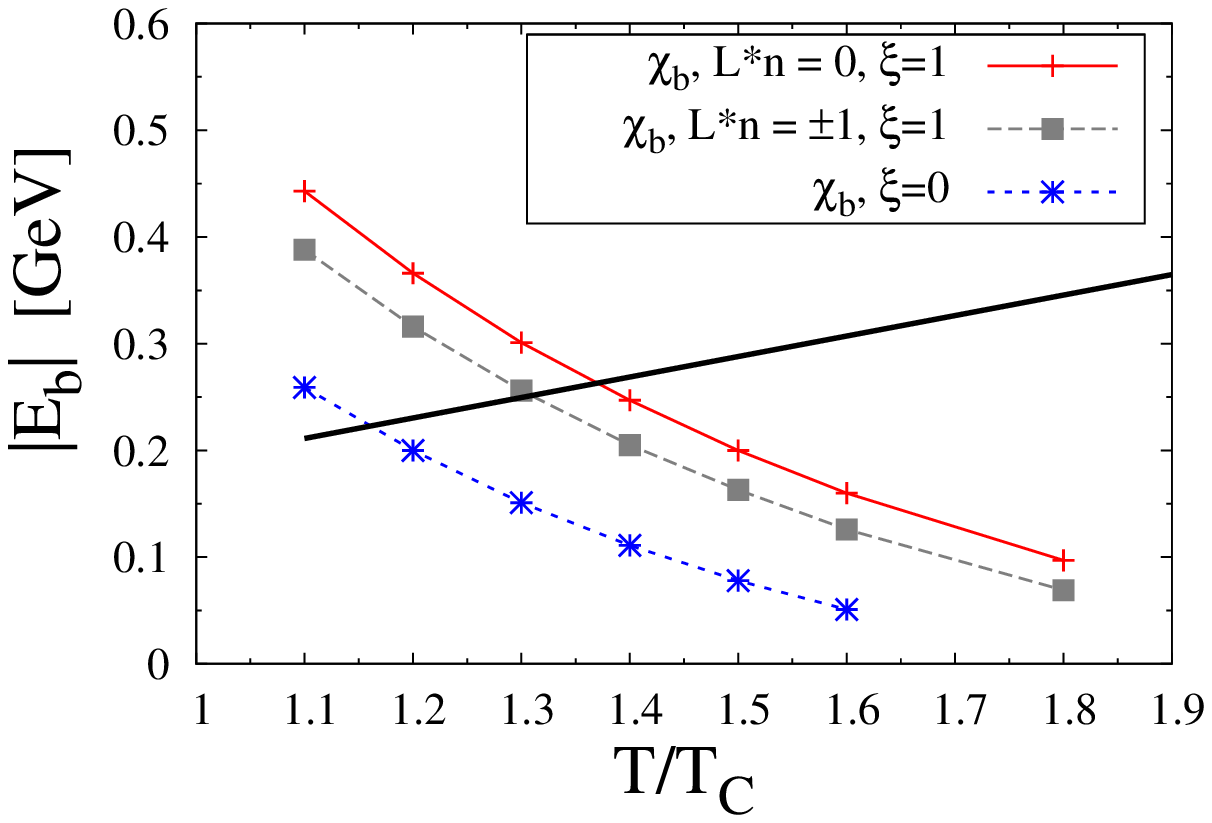}
\end{center}
\caption[a]{Left: Spectral function of pseudo-scalar bottomonium
  obtained from a potential model~\cite{Mocsy:2007yj} with
  $V_\infty(T) \sim 1/{T}$. In this calculation, the width of the
  peaks has been put in by hand for numerical stability. Note the
  logarithmic scale.  Right: Binding energy of $P$-state bottomonium
  ($L=1$) as a function of temperature for an isotropic ($\xi=0$) and
  a moderately anisotropic ($\xi=1$) QGP~\cite{Dumitru:2009ni}. The
  solid increasing line shows $T$ itself; note that $T_c=192$~MeV was
  assumed here to normalize the abscissa.}
\label{fig:SpecFuncUps}
\end{figure}
For semi-realistic estimates of the binding energies of charmonium and
bottomonium states one needs to add the linear confining potential
$\sim \sigma r$, where $\sigma\simeq1$~GeV/fm is the SU(3) string
tension. Its temperature dependence to be used in the real-time
formalism must presently be modelled. In the phenomenologically
relevant range $T/T_c = 1-3$ the ``interaction measure'' $(e-3p)/T^4$
in QCD is large. The potential at infinite separation is thus
sometimes modelled as~\cite{Mocsy:2007yj} $V_\infty(T) \simeq
2{a}/{T}$ with $a\approx 0.08$~GeV$^2$ a constant of dimension
two. Such a model has in fact been proposed long ago in ref.~\cite{KMS}
\be \label{KMSpot}
V(r) = \left[ -\frac{\alpha_s C_F}{r} + 2\frac{\sigma}{m_D} 
(e^{\hat{r}}-1) - \sigma r \right] \, e^{-\hat{r}}~,
\ee
which also provides a smooth interpolation to short distances. In this
model the temperature dependence of the binding energy $E_{\rm bind}=
\langle \Psi \left| \hat{H} - V_\infty \right| \Psi\rangle - 2m_Q$ of
small bound states such as the $\Upsilon$ actually turns out to
originate mostly from $V_\infty(T)$~\cite{Dumitru:2009ni}. This is
confirmed by the spectral function of pseudo-scalar bottomonium
published in ref.~\cite{Mocsy:2007yj} which has been replotted on a
logarithmic scale in fig.~\ref{fig:SpecFuncUps}: while the ground
state peak shows little temperature dependence, the continuum
threshold decreases rapidly as $T$ increases. Similar spectral
functions have been shown in ref.~\cite{Riek:2010fk}.

The anisotropy or viscosity affects the excited states more strongly
than the compact $1S$ bottomonium state. In this model the binding
energy at $T/T_c=1.1$, for example, increases by about
50\%. Alternatively, if a ``dissolution'' temperature is defined from
$|E_b|=T_{\rm dis}$ then this increases from $\simeq1.15T_c$ when
$\xi=0$ to about $1.35T_c$ when $\xi=1$. Thus, excited states should
melt less easily if the QGP exhibits an anisotropy during the early
stages of the expansion. Lastly, there is a splitting of the states
with ${\bf L}\cdot {\bf n}=0$ and ${\bf L}\cdot {\bf n}= \pm 1$,
respectively, since rotational symmetry is broken. Understanding the
properties of excited states of the $\Upsilon$ is important for
phenomenology because they contribute through feed down to the yield
of the $1S$ state~\cite{Liu:2010ej}.

\subsection{Imaginary part of the quarkonium potential}
At finite temperature, the quarkonium potential also acquires an
imaginary part~\cite{Laine:2006ns} at order $g^2C_F$ due to Landau
damping of the exchanged nearly static gluon. It can again be obtained
from the Fourier transform of the HTL-resummed real-time propagator
(the ``physical'' component of the Schwinger-Keldysh representation)
for static $A_0$ fields. Taking the imaginary part corresponds to
cutting open one of the hard thermal loops of the HTL propagator and
can be viewed, microscopically, as the scattering of the space-like
exchanged gluon off a thermal gluon~\cite{Laine:2006ns}:
$g+(Q\bar{Q})\to g+Q+\bar{Q}$. To order $\xi$ and in the leading $\log
1/\hat{r}$ approximation ${\rm{Im}} \, V$ is given
by~\cite{Dumitru:2009fy}
\begin{eqnarray}
{\rm Im} \, V({\bf{r}}) =-\frac{g^2 C_F
T}{4\pi}\,\hat{r}^2\log\frac{1}{
\hat{r}} \, \left(\frac{1}{3}-\xi\frac{3-\cos(2\theta)}{20} \right)~.
\end{eqnarray}
This corresponds to a decay width for a Coulomb ground state of
\begin{eqnarray}
\Gamma &=& \frac{16\pi T}{g^2
C_F}\frac{m_D^2}{M^2_Q} \left(1-\frac{\xi}{2}\right)
\log\frac{g^2C_FM_Q}{4\pi m_D}~.
\end{eqnarray}
For $\xi=0$ one can in fact evaluate the matrix element of ${\rm{Im}}
\, V$ between Coulomb wave functions without resorting to the $\log
1/\hat{r} \gg1$ approximation,
\be \label{GammaXi0}
\Gamma = \frac{T}{\alpha_s C_F} \frac{m_D^2}{M_Q^2}
\frac{1-(2-\kappa^2)^2 + 4\log\frac{1}{\kappa} }{(1-\kappa^2)^3} ~~~,~~~
\kappa = \frac{1}{\alpha_s C_F} \frac{m_D}{M_Q}~.
\ee

\begin{figure}[hbt]
\begin{center}
\includegraphics[width=3.8cm]{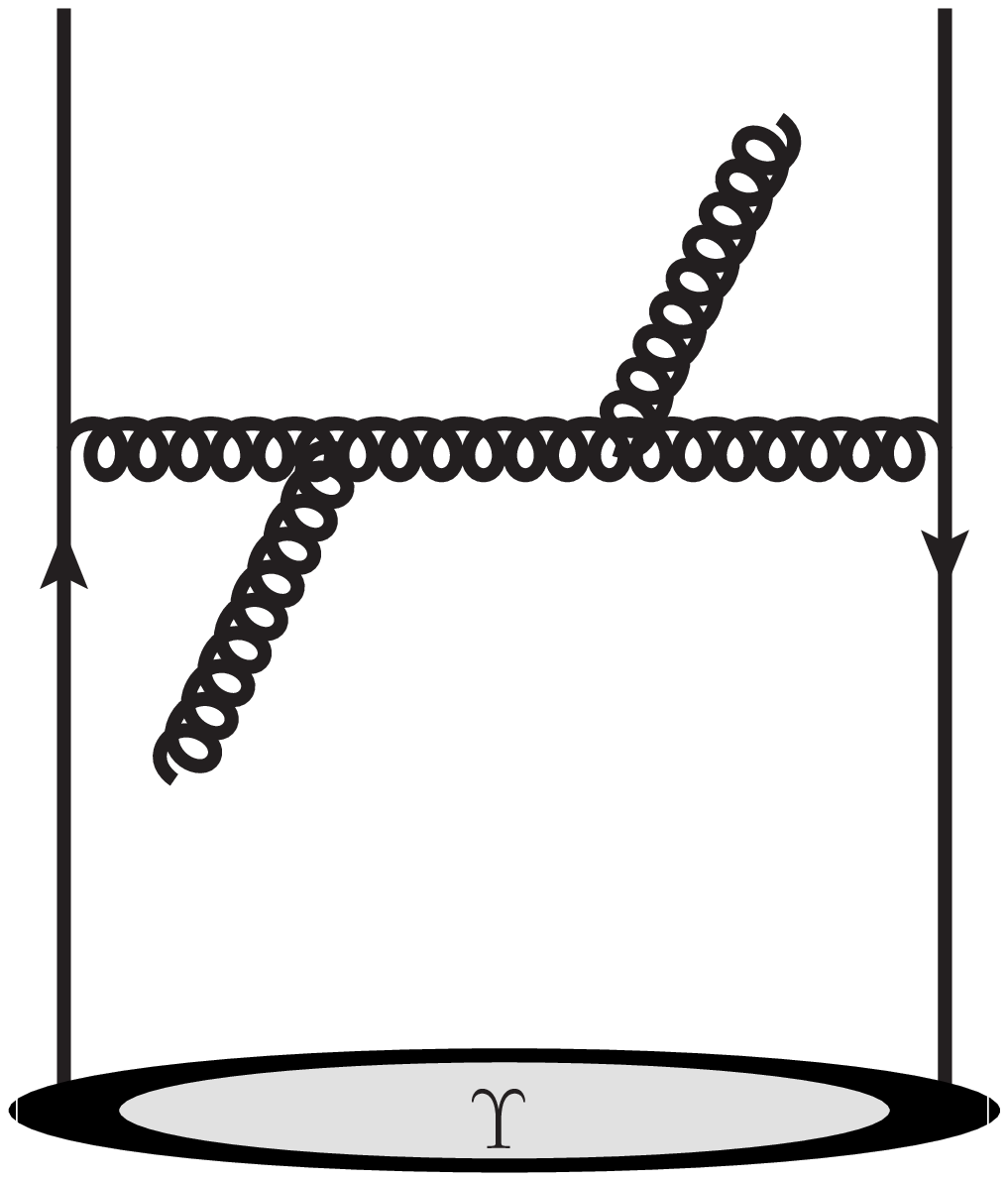}
\includegraphics[width=7cm]{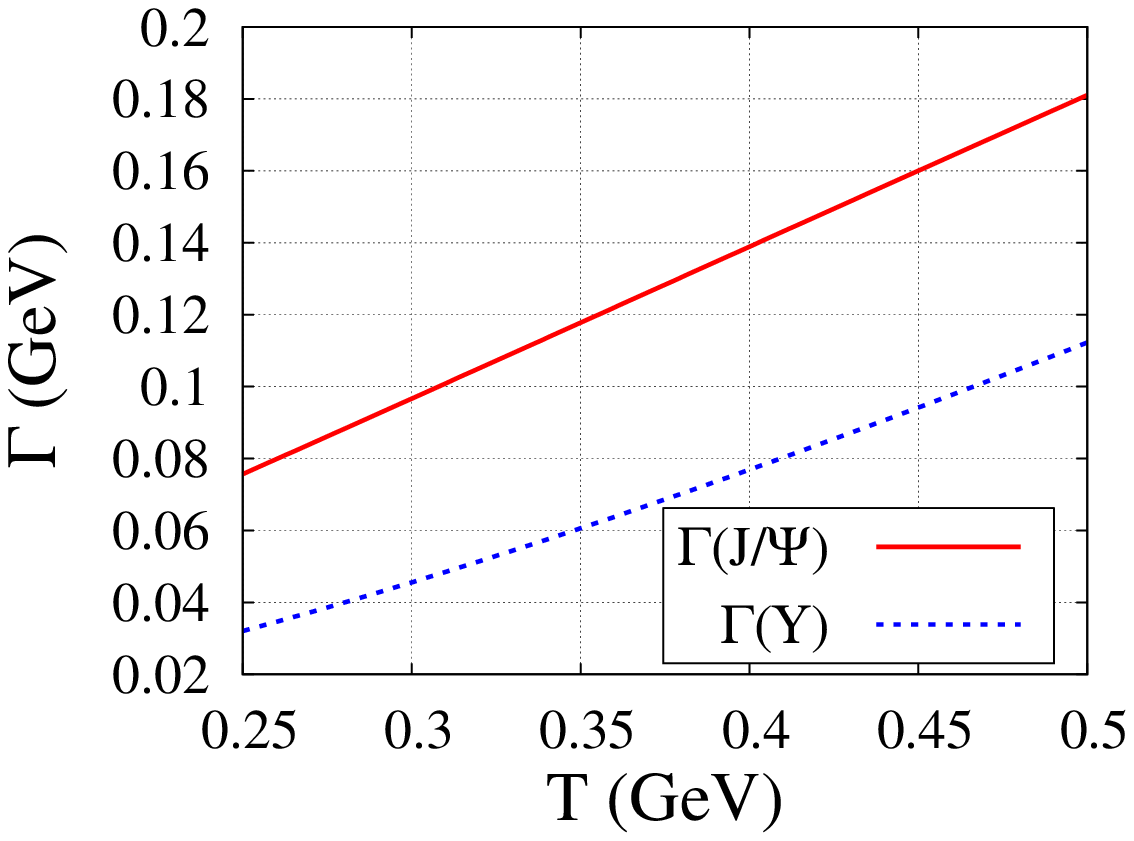}
\end{center}
\caption[a]{Left: Imaginary part of the real-time propagator due to
  Landau damping.  Right: Thermal decay width as a function of
  temperature for an isotropic medium ($\xi=0$).}
\label{fig:Gamma}
\end{figure}
The width obtained from~(\ref{GammaXi0}) is shown in
fig.~\ref{fig:Gamma}. For temperatures accessible to the RHIC and LHC
colliders, $\Gamma_\Upsilon$ is on the order of 20~MeV --
100~MeV. This can be compared to the $\Upsilon\to e^+ e^-$ decay width
in vacuum which arises from the annihilation of the $b$, $\bar{b}$
quarks into di-electrons\footnote{It is therefore proportional to the
  square of the quarkonium wave function at the origin, similar to the
  hadronic decay into three gluons discussed by Appelquist and
  Politzer~\cite{AP75} but unlike the
  thermal QCD width discussed above.}: $\Gamma_{\Upsilon\to e^+ e^-} =
1.34$~keV~\cite{PDG}. Because the electromagnetic decay width is much
smaller than the inverse lifetime of the QGP the actual $\Upsilon$
peak observed in the di-lepton invariant mass distribution would not
be broadened as compared to vacuum.

Nevertheless, $\Upsilon$ states which have been broken up into
individual $b$ and $\bar{b}$ quarks reduce the yield of di-leptons in
the peak. One of the contributions to this process is $g+(Q\bar{Q})\to
Q+\bar{Q}$ dissociation~\cite{BP} by a thermal
gluon~\cite{Liu:2010ej,Grandchamp:2005yw}. However, it has been found
that this process leads to rather small $\Upsilon$ dissociation
rates. Indeed, ref.~\cite{Grandchamp:2005yw} argues that in the limit
of loose binding, $|E_b|\ll T$ due to strong screening of the
attractive interaction, it becomes more efficient to scatter a thermal
gluon off one of the quasi-free $b$-quarks thereby breaking up the
bound state. The width~(\ref{GammaXi0}) obtained from the imaginary
part of the HTL propagator indicates that damping of the exchanged
gluon in the heat bath also provides a large contribution to the
thermal $\Upsilon\to b+\bar{b}$ rate. 

At low temperature, once pions have formed, a non-zero (but
exponentially small) width emerges due to $\pi + \Upsilon\to
B+\bar{B}$ corresponding to tunneling of the $b$, $\bar{b}$ quarks in
the background field of the pion~\cite{Kharzeev:1995ju}.

It is certainly interesting to compare to a strongly coupled
theory. Using the gauge-gravity duality, the static potential (or
Wilson loop)~\cite{Maldacena:1998im} and thermal effects at short
distances~\cite{Rey:1998bq} have been computed in ${\cal N}=4$
supersymmetric Yang-Mills at large (but finite~\cite{Gubser:2006qh})
t' Hooft coupling $\lambda=g^2 N$ and $N\to\infty$. At $T=0$,
\begin{equation}
V_{Q\bar{Q}}(r)= -\frac{4\pi^2}{\Gamma(1/4)^4} \frac{\sqrt{\lambda}}{r}~.
\label{maldacenaprop}
\end{equation}
The $\sim 1/r$ behavior follows from conformal invariance of the
theory. Also, the potential is non-analytic in $\lambda$. Clearly, the
coupling should not be very large or else the properties of the
resulting bound states are qualitatively different from the $\Upsilon$
etc.\ states of QCD (numerically, $4\pi^2/\Gamma(1/4)^4 \approx
0.23$).

At finite temperature, the potential develops an imaginary part when
the string dangling in the fifth dimension (with end points on
our brane) approaches the black hole horizon which sets the
temperature scale and ``melts''~\cite{Albacete:2008dz}. Fluctuations
near the ``bottom'' of the string should generate such an imaginary
part of the Nambu-Goto action at even lower temperatures
already~\cite{Noronha:2009da},
\begin{eqnarray}
\Gamma_{Q\bar{Q}} &=& -\langle \psi|
\mathrm{Im}~V_{Q\bar{Q}} |\psi\rangle \simeq
\frac{\pi\sqrt{\lambda}}{48\sqrt{2}}\frac{b}{a_0}
 \left[45\left(\frac{a_0T}{b}\right)^4 -2\right]~,
\label{AdSwidth}
\end{eqnarray} 
where $|\psi\rangle$ denotes the unperturbed Coulomb ground state wave
function, $a_0=\Gamma(1/4)^4/2\pi^2\sqrt{\lambda}\,m_Q$ is the ``Bohr
radius'' for the Maldacena potential~(\ref{maldacenaprop}) and
$b=2\Gamma(3/4)/\sqrt{\pi}\,\Gamma(1/4)\approx 0.38$ is a numerical
constant. Here the width decreases with the quark mass and with the t'
Hooft coupling approximately as $\Gamma_{Q\bar{Q}}\sim 1/\lambda\,
m_Q^3$; it increases rapidly with the temperature, $\sim T^4$. For
$m_Q=4.7$~GeV, $T=0.3$~GeV, $\sqrt\lambda=3$ one finds
$\Gamma_\Upsilon \simeq 50$~MeV. The $(Q\bar{Q})\to (Q\bar{q})\,
(\bar{Q}q)$ breakup due to string splitting at has been considered in
ref.~\cite{Cotrone:2005fr}.

The suppression of the yield of di-leptons from $\Upsilon(1S)$ decays
in the final state should be
significant~\cite{Grandchamp:2005yw,Noronha:2009da}. Neglecting
``regeneration'' of bound states from $b$ and $\bar{b}$ quarks in the
medium, the number of $\Upsilon$ mesons in the plasma which have not
decayed into unbound $b$ and $\bar{b}$ quarks up to time $\tau$ after the
collision is
\begin{eqnarray}
N(t) &\simeq& N_0 \exp\left(-\int^\tau_{\tau_0} dt\, 
\Gamma_\Upsilon(t)\right)~.
\end{eqnarray}
This solution assumes that $\Gamma_\Upsilon(T(t))$ is a slowly varying
function of time. The initial number of $\Upsilon$ states may be
estimated from the multiplicity in p+p collisions times the number of
binary collisions at a given impact parameter: $N_0\simeq N_{\rm coll}
N_{pp}^\Upsilon$. Thus, the ``nuclear modification factor'' $R_{AA}$
for the process $\Upsilon\to \ell^+ \ell^-$ is approximately given by
$R_{AA}(\Upsilon\to \ell^+ \ell^-) \simeq \exp(-\bar\Gamma_\Upsilon \,
\tau)$, where $\bar\Gamma$ denotes a suitable average of $\Gamma(T)$ over
the lifetime of the quark-gluon plasma. Due to the strong temperature
dependence of the width, this average is dominated by the early
stage. Experimental measurements of $R_{AA}(\Upsilon\to \ell^+
\ell^-)$ at RHIC appear to indicate a suppression~\cite{data} but so
far it has not been possible to disentangle various $\Upsilon$ states.

\section{The static gluon propagator in the ``semi-QGP''}
This section is about the HTL propagator in Euclidean time for
temperatures just above deconfinement. Hidaka and Pisarski suggested
that for $T$ not far above $T_c$ hard modes in the QCD plasma may
still be weakly coupled but propagating in a non-perturbative
background $A_0$ field~\cite{HP}. The expectation value of the
Polyakov loop is parameterized as
\be
{\bf L} = \exp \left(-i \frac{Q}{T}\right) ~~~,~~~ Q\equiv g A_0~~~,~~~
\ell = \frac{1}{N} {\rm tr}~{\bf L}~.
\ee
\begin{figure}[hbt]
\begin{center}
\includegraphics[width=6.5cm]{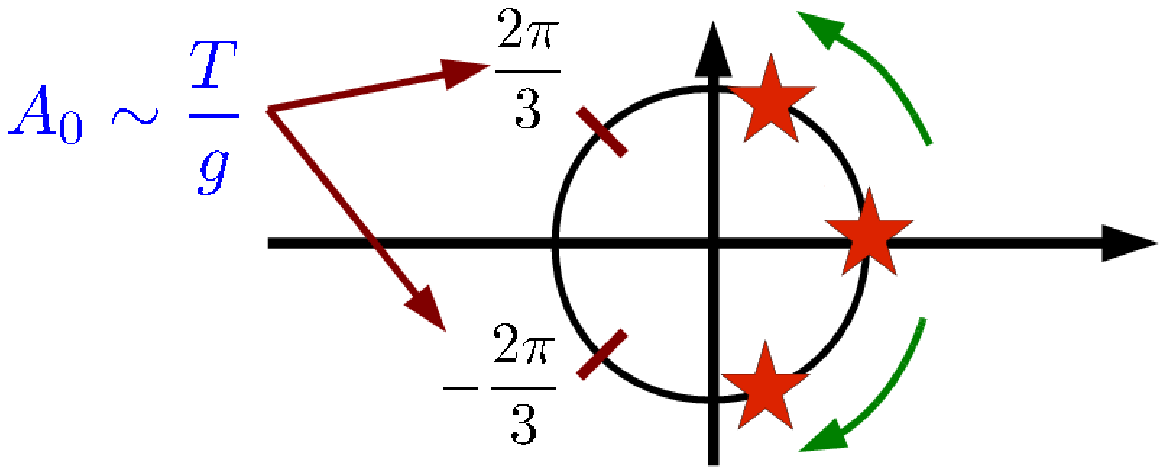}
\includegraphics[width=6.5cm]{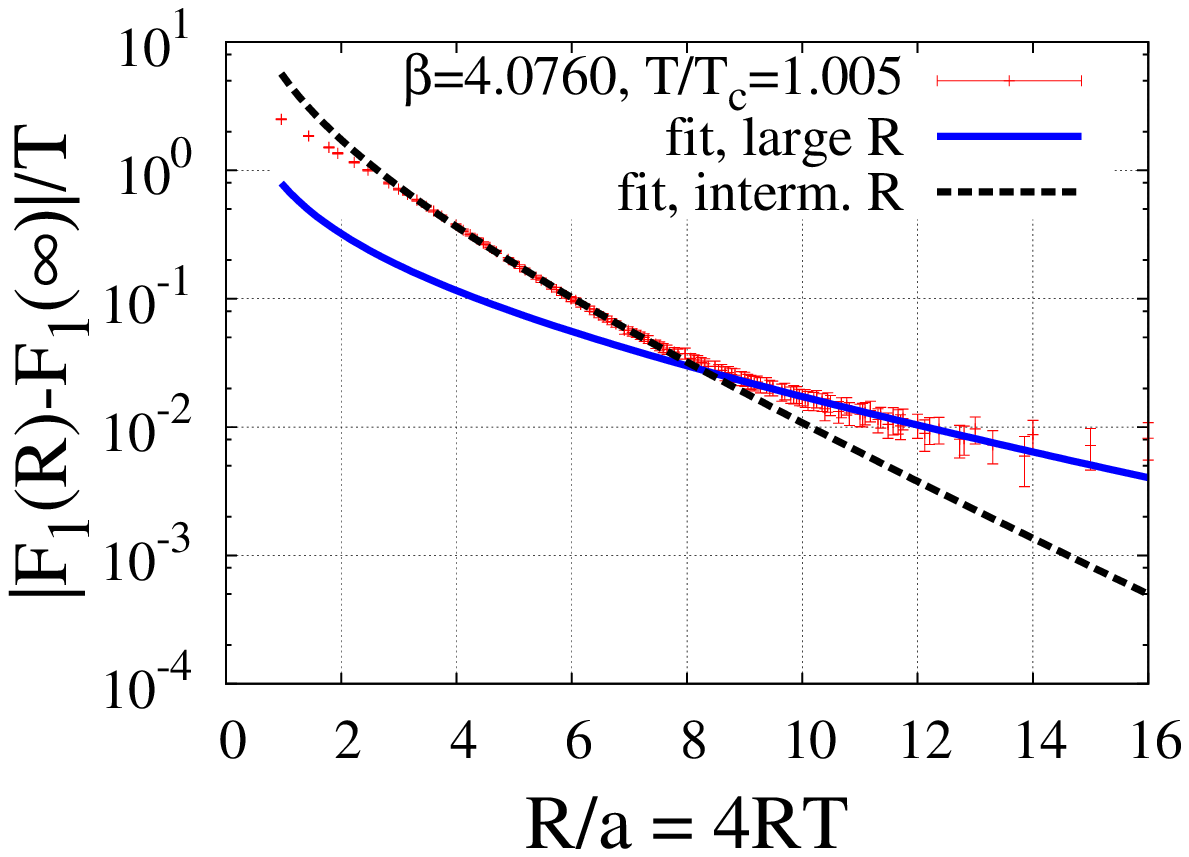}
\end{center}
\caption[a]{Left: Illustration of ``repulsion'' of the {\em eigenvalues} of
  the Polyakov loop for $N=3$ colors. Right: Singlet free energy
  versus distance $R$, data from~\cite{F1_Olaf}.}
\label{fig:EVloop}
\end{figure}
At high $T$ the {\em eigenvalues} of the matrix ${\bf L}$ approach 0
and so its normalized trace $\ell\to1$. On the other hand, as $T\to
T_c$, the {\em eigenvalues} ``repell'' and eventually, at $T_c-0$,
distribute uniformly over a circle such that $\ell=0$. Note that
$\ell(T_c+0) \approx 0.4$ in SU(3) Yang-Mills~\cite{RenLoop}, far from
unity, and hence that $A_0\sim T/g$ is non-perturbatively large. It is
in this sense that the ``semi-QGP'' just above $T_c$ is a weakly
coupled but non-perturbative phase.

Gluons which couple to the background field acquire an additional
``mass''; the static propagator in the semi-QGP (summed over colors)
becomes, schematically\footnote{Eq.~(\ref{eq:Prop_sQGP}) is a
  simplified schematic expression. The complete expression shall be
  given in ref.~\cite{DGHKP}.}
\be \label{eq:Prop_sQGP}
\frac{1}{N}\langle {\rm tr}~{\bf L}^\dagger(R) \, {\bf L}(0)\rangle \sim 
g^2 C_F
\int \frac{d^3p}{(2\pi)^3} e^{i px} \sum\limits_{kl} \frac{1}
{p^2 + m_D^2 + (Q^l - Q^k)^2} ~.
\ee
The $N$ diagonal gluons are screened only over large distances of
order $1/m_D$, where $m_D^2={\cal O}(g^2 T^2 N)$ is the usual Debye
mass. The $N^2-N$ off-diagonal gluons corresponding to $l\ne k$,
however, acquire a mass of order $T$ when $A_0^a-A_0^b\sim T/g$. 
The heavy-quark free energy $F_{Q\bar{Q}}(R)/T$ defined from the
correlator of Polyakov loops (which may not necessarily be identified
with the potential in real time) is expected to show the following
qualitative behavior (assuming, for simplicity, large $N$): 
\bea
RT\ll 1: ~~~& & F_{Q\bar{Q}}(R) \sim - \frac{g^2 N^2}{N R} \sim
\frac{g^2 N}{R} \\
1\ll RT\ll \frac{1}{g}:~~~& & F_{Q\bar{Q}}(R) \sim - \frac{g^2 N}{N R} \sim
\frac{g^2}{R} \label{intDC}\\
\frac{1}{g} \ll RT:~~~ & & F_{Q\bar{Q}}(R) \sim - \frac{g^2 N}{N R}\,
e^{-m_D R} \sim \frac{g^2}{R}\, e^{-m_D R}~.  \label{longDC}
\eea
In the intermediate region~(\ref{intDC}) the $N^2-N$ heavy
off-diagonal gluons have decoupled and $F_{Q\bar{Q}}$ is suppressed by
$1/N$ as compared to its behavior at short distances. In
fig.~\ref{fig:EVloop} we show the singlet free energy for SU(3)
Yang-Mills just above $T_c$ obtained numerically on $32^3\times4$
lattices~\cite{F1_Olaf}. Indeed, it does appear to show a two-slope
behavior with a transition at $RT\simeq2$. On the other hand, near
$T_c$ the large distance behavior could be affected by finite volume
artifacts~\cite{OKpriv}; simulations on larger lattices may hopefully
become available in the future. If the behavior seen in the present
data is confirmed then this may indicate that just above $T_c$ the
properties of large, excited quarkonium states may be modified as
compared to predictions from a standard potential model such
as~(\ref{KMSpot}).

\section*{Acknowledgements}
I thank all of my collaborators for their input and
Prof.\ K.~Yazaki for useful questions about the quarkonium width
during the HESI2010 meeting.  I am indebted to the organizers of the
Yukawa International Program for Quark-Hadron Sciences at Yukawa
Institute for Theoretical Physics, Kyoto University, for their
hospitality and for the invitation to present this work at the ``High
Energy Strong Interactions 2010'' symposium. I gratefully acknowledge
support by the DOE Office of Nuclear Physics through Grant
No.\ DE-FG02-09ER41620, by the Dean's office of Weissman School of
Arts and Sciences, and by Yukawa International Program for
Quark-Hadron Sciences.

%

\end{document}